# Magnetization reversal in the anisotropy-dominated regime using time-dependent magnetic fields

K. Rivkin[1]  and  J. B. Ketterson[1,2,3]

1. Department of Physics and Astronomy, Northwestern University
Evanston IL, 60201

2. Department of Electrical and Computer Engineering, Northwestern University
Evanston IL, 60201

3. Materials Research Center, Northwestern University
Evanston IL, 60201

## Abstract

We study magnetization reversal using various r.f. magnetic pulses. We show numerically that switching is possible with simple sinusoidal pulses; however the optimum approach is to use a frequency-swept (chirped) r.f. magnetic pulse, the shape of which can be derived analytically. Switching times of the order of nanoseconds can be achieved with relatively small r.f. fields, independent of the anisotropy's strength.

Among the underlying physics issues associated with magnetic memory devices, perhaps the most subtle process is that by which the magnetization direction is reversed (switched). As far as we are aware, Thirion et al[1] were the first to show, experimentally, that in the presence of magnetic anisotropy, reversing the magnetic field while simultaneously applying an r.f. field at right angles to the applied d.c. field can significantly lower the threshold for switching. These authors also concluded that the most effective frequency for small amplitude oscillations corresponded to the uniform-mode fmr frequency. Whether this result can be exploited commercially is an open question.

Prior to the work cited, reversal via r.f. fields had been studied theoretically.[2,3] Bauer et al[4] showed that *in the absence* of a constant, $H_0$, field, a d.c. pulse (one for which the field remains positive throughout the pulse) can produce switching in materials with uniaxial anisotropy. However the switching times reported were quite long, of the order of 500ns for typical conditions. In this paper we explore constant-frequency as well as swept (chirped) r.f. pulse profiles in order to better optimize the switching time.

We identify two different regimes in which to consider magnetization reversal via r.f. pulses, depending whether $H_0$ is greater than or less than the effective anisotropy field. In the first case one might excite large amplitude uniform-precession at the ferromagnetic resonance (fmr) frequency, analogous to what is done in nuclear magnetic resonance experiments.[5] This approach has not been particularly effective in ferromagnets due the onset of the so-called *Suhl instability*[6] which arises from the non-linear excitation of non-uniform spin modes; this phenomenon ultimately leads to spatial decoherance and, for sufficiently long times and high values of the applied r.f. field, chaos.[7]  However, as shown in reference[8], this instability can be greatly suppressed by choosing favorable values of the $H_0$ field and sample size.  Furthermore, in most cases, this instability becomes irrelevant for objects smaller than 300nm.

Most magnetic memories utilize, in one way or another, magnetic anisotropy to store information. Even though the application an r.f. pulse can lower the d.c. field required to induce switching, this approach is not applied in practice and clearly the engineering is simplest if d.c. fields can be avoided.

Here we will limit our numerical studies to the case where $H_0 = 0$.  As material parameters we choose a bismuth-substituted YIG sphere having a saturation magnetization $M_s = 120\,Oe$ and a damping coefficient $\beta = 0.001$; the anisotropy coefficient $K = 7.3 \cdot 10^4$ erg/cm corresponds to an effective anisotropy field of 608 Oe or, equivalently, a small-amplitude precession frequency $\omega_K / 2\pi = 1.69\,GHz$.  We will assume that the sphere can be approximated by a *single spin*; based on our simulations with ensembles, this turns out to be reasonably accurate if the diameter is less than 300 nm.  We further assume that the z-axis is the easy axis with the magnetization initially along the positive direction.

At t = 0 we apply a circularly polarized r.f. field in the x-y plane having the form

$$
\begin{aligned}
H_x^{rf} &= H_1 \cos\big(\phi(t)\big) \\
H_y^{rf} &= H_1 \sin\big(\phi(t)\big)
\end{aligned}
\qquad (1)
$$

where in our simulations the amplitude $H_1$ will be constant during the application of the pulse and zero otherwise.  For a constant applied frequency $\phi(t) = \phi_0 + \omega t$; for a swept pulse we define the *instantaneous* frequency as $\omega(t) = \dfrac{d\phi}{dt}$.  We then solve the Landau-Lifshitz[9] equation

$$\frac{d\mathbf{m}}{dt} = -\gamma \mathbf{m} \times \mathbf{h}^{\text{total}} - \frac{\beta\gamma}{M_s} \mathbf{m} \times \left( \mathbf{m} \times \mathbf{h}^{\text{total}} \right) \qquad (2)$$

using a $4^{\text{th}}$ order Runge-Kutta algorithm; here $\mathbf{m}$ is the magnetization and $\mathbf{h}^{\text{total}}$ is the sum of the applied r. f. field and the anisotropy field.

We first examine the case of a constant applied frequency and ask which frequencies produce the maximal rotation of the magnetization.  Figure 1 shows $M_z$ as a function of $\omega / 2\pi$ for two different values of r.f. field – 50 and 100 Oe.  For our single-spin model, switching will occur following a rotation by an angle greater than $\pi / 2$; the moment must then subsequently relax to the negative z direction.  Note that relatively high r.f. fields are required to produce switching.  On the other hand rotation induced by a d.c. pulse applied perpendicular to the z axis requires still larger fields.

We note that the frequency associated with the anisotropy field, $\omega_K(\theta)$, is *angle dependent*, going to zero at $\theta = \pi / 2$.  For this reason we find that $\omega_K(0)$ is not the optimal frequency for switching; the optimal frequency turns out to be about half $\omega_K(0)$.  Moreover switching is possible only within a set of narrow frequency ranges.  Switching with a constant frequency has another disadvantage – while it takes only a few nanoseconds for $M_z$ to become negative, full switching ($\theta = \pi$) takes about 500ns to complete, the time dependence of $M_z$ being mostly oscillatory (implying the necessity of a significant time separation between "writing" and reliably reading or rewriting the magnetic state). Because of this the optimal switching strategy is to use a short r.f. pulse with a carefully chosen frequency and width; one can then reach values of $M_z/M_s$ as low as $-0.35$ before the system starts to oscillate back towards the positive values of $M_z$. At this point we turn off the r.f. field, and wait for the damping to relax the system towards $M_z/M_s = -1$. Figure 2 shows the times at which $M_z/M_s$ goes through 0 as a function of frequency. The narrow regions between two neighboring curves define intervals for which the system has $M_z/M_s < 0$, and if the r.f. field is switched off during one of these intervals, the system will damp down to $M_z/M_s = -1$. Unfortunately, for most of the time $M_z$ is positive, and therefore the pulses lengths must be quite precise.

As noted above, $\omega_K(\theta)$ is angular dependent and, since $\theta = \theta(t)$, this frequency is consequently time dependent. Because of this, it is natural to apply a pulse for which the frequency is *time dependent*, a so-called *chirped* pulse. Initially we attempted to numerically calculate the frequency $\omega(t)$ which yields the fastest possible switching. The results of this calculation are presented on the Figure 3. However it was then realized that this optimal switching can be expressed as an analytical solution of Landau-Lifshitz equation (in the most general case a dc field $H_0$ is also applied along z axis, however its presence is unnecessary):

$$\dot{M}_x = -\gamma \left( M_y \left( K \frac{M_z}{M_s^2} + H_0 \right) - M_z H_1 \sin(\varphi(t)) \right)$$

$$\dot{M}_y = -\gamma \left( M_z H_1 \cos(\varphi(t)) - M_x \left( K \frac{M_z}{M_s^2} + H_0 \right) \right) \qquad (3)$$

$$\dot{M}_z = -\gamma \left( M_x H_1 \sin(\varphi(t)) - M_y H_1 \cos(\varphi(t)) \right);$$

in order for switching to be optimal, $\omega(t)$ should always be equal to the resonant frequency:

$$\omega(t) \equiv \frac{d\varphi(t)}{dt}$$
$$= \gamma \frac{K}{M_s^2} M_z + \gamma H_0. \tag{4}$$

Integrating Eq. (4) we can obtain the expression for $\varphi(t)$ (the angle that the r.f. field forms with x axis), and thereby solve Eq.(3):

$$M_x = M_s \sin(\gamma H_1 t) \sin(\varphi(t))$$
$$M_y = -M_s \sin(\gamma H_1 t) \cos(\varphi(t)) \tag{5}$$
$$M_z = M_s \cos(\gamma H_1 t)$$

$$\varphi(t) = K \frac{\sin(\gamma H_1 t)}{H_1 M_s} + \gamma H_0 t \tag{6}$$

As can be seen in Figure 3, $\omega(t)$ always corresponds to the anisotropy frequency, $\omega_K(\theta)$; initially it is equal to $\omega_K(0)$, then goes through zero when $M_z(t) = 0$, and is equal to $-\omega_K(0)$ when $M_z(t) = -M_s$.

The stability of this chirped-frequency switching was tested by applying a random, uniformly-distributed, noise to $\varphi(t)$: the switching still occurs for values of the noise as high as 25% of $\varphi$.

If for technical reasons we are constrained to r.f. fields applied only along one of the coordinate axis (for example the x axis), it can be that shown that a series of square (flat-topped) pulses of equal magnitude, alternating sign, and varying width produces nearly optimal switching; such a waveform might be synthesized digitally in practice.

The associated pulse widths are chosen so as to maximizes the decrease in $M_z$. The rate of change of $M_z$ due to the applied r.f. field is given by

$$\frac{dM_z}{dt} = \gamma M_y H_x^{rf} ; \tag{7}$$

hence if we wish to minimize this term with the chosen waveform, the sign of $H_x^{rf}$ should be opposite to that of $M_y$. In figure 4 we present the resulting pulse sequence and the behavior of $M_z / M_s$ vs. time for the case where the amplitude of r.f. field $H_1 = 25\,Oe$.

In conclusion, we have established that efficient switching of the magnetization direction of a spin precessing under the joint influence of an anisotropy field and an external r.f. field can be achieved in the absence of an applied d.c. field. We find the following behaviors:

i)   The optimal frequency for switching with a *non-chirped* r.f. pulse is approximately half the small-amplitude precession frequency associated with the anisotropy energy; switching is possible only for r. f. pulses with amplitudes above some threshold value. Due to the oscillatory nature of such switching, the pulse length must be chosen precisely, terminating it before the system goes back to positive value of $M_z$; the remainder of the switching occurs via damping.

ii)  Switching with a chirped r.f. pulse is orders of magnitude superior to case i). In this case the switching can be accomplished using pulses of arbitrarily low power (in the absence of damping), albeit with a switching time inversely proportional to the value of the applied r.f. field. Switching times do not depend on the value of anisotropy and saturation magnetization. Not only does the switching occur much faster than for the case of non-chirped pulses, it is also more stable with respect to the random changes in the pulse parameters.

The program used for these simulations is available for the public use at www.rkmag.com. This work was supported by the National Science Foundation under grant ESC-02-24210.

**Figure captions:**

**Figure 1. Minimal value of $M_z$ as a function of frequency for two values of the r.f. field** $H_1 = 100\,\mathrm{Oe}$ **and** $H_1 = 50\,\mathrm{Oe}.$

**Figure 2. Time when $M_z$ is equal to 0 as a function of frequency for r.f. field**
$H_1 = 100\,\mathrm{Oe}.$

**Figure 3. Values of phase, $M_z$, $M_x$ and frequency as a function of time for**
$H_1 = 100\,\mathrm{Oe}.$

**Figure 4. Values of $M_z$ and the r.f. field as a function of time for** $H_1 = 25\,\mathrm{Oe}$ **applied only along the x direction.**

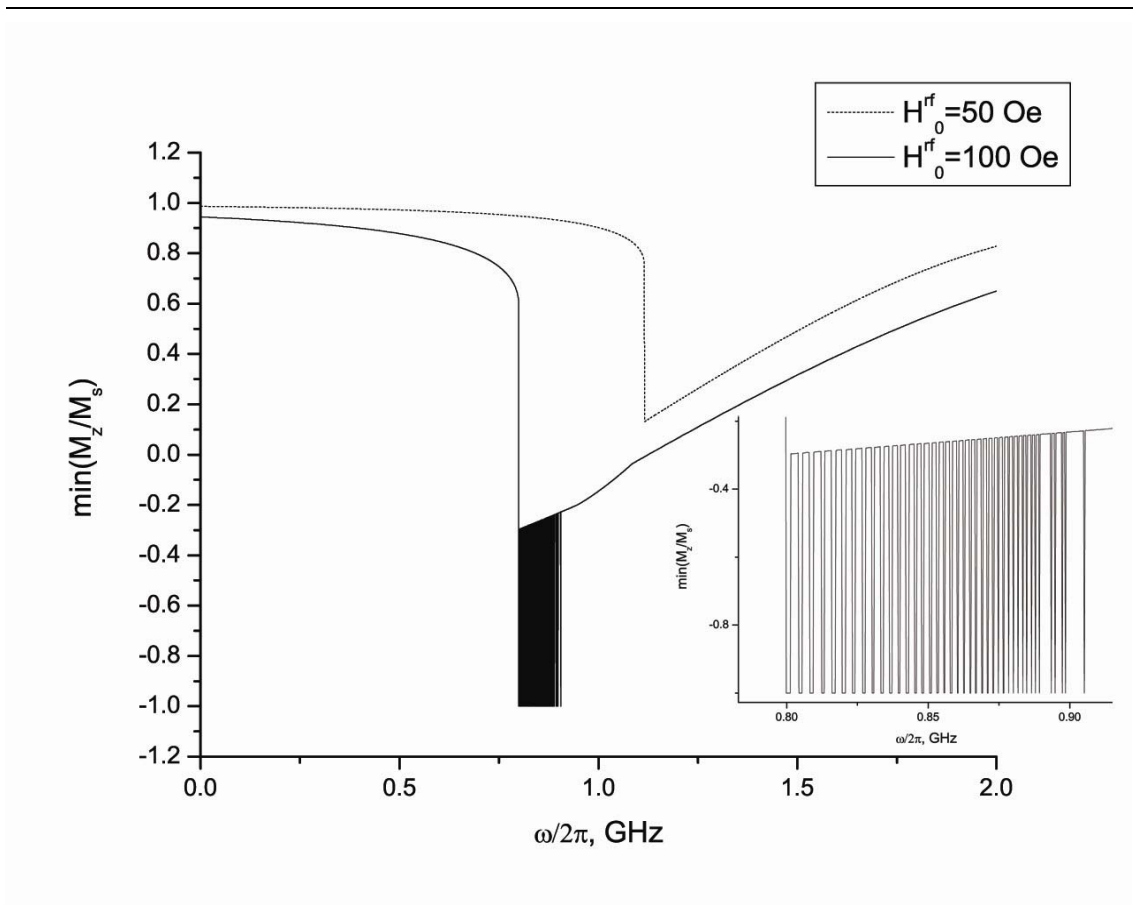

**Figure 1. Minimal value of $M_z$ as a function of frequency for two values of r.f. field** $H_1 = 100\,\text{Oe}$ **and** $H_1 = 50\,\text{Oe}$.

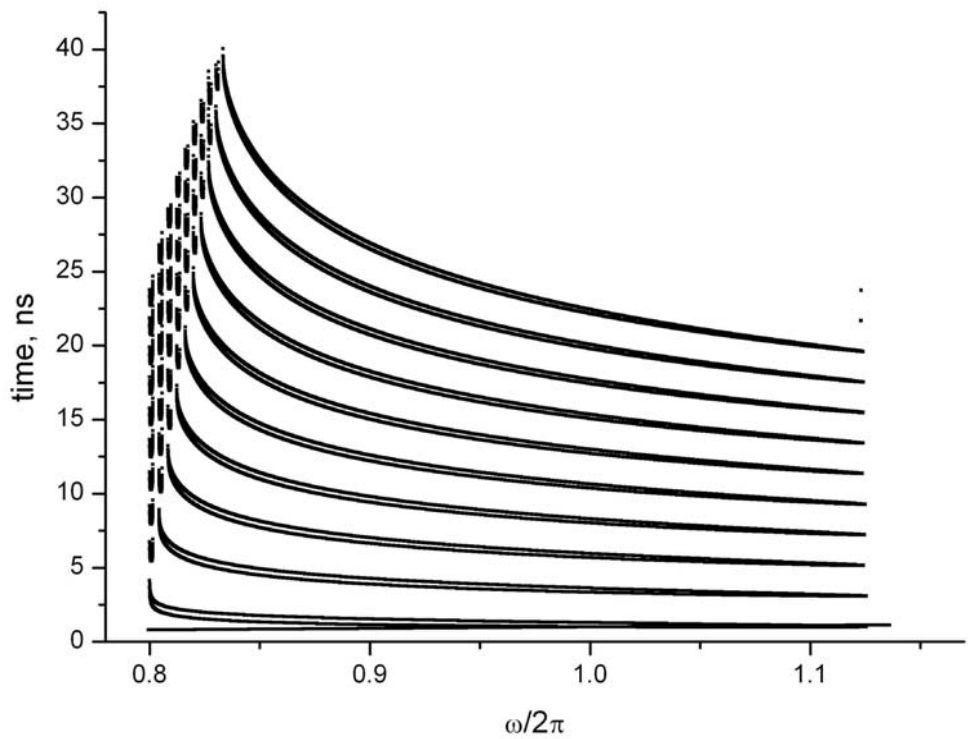

**Figure 2. Time when M$_z$ is equal to 0 as a function of frequency for r.f. field**

$H_1 = 100 \, \text{Oe}.$

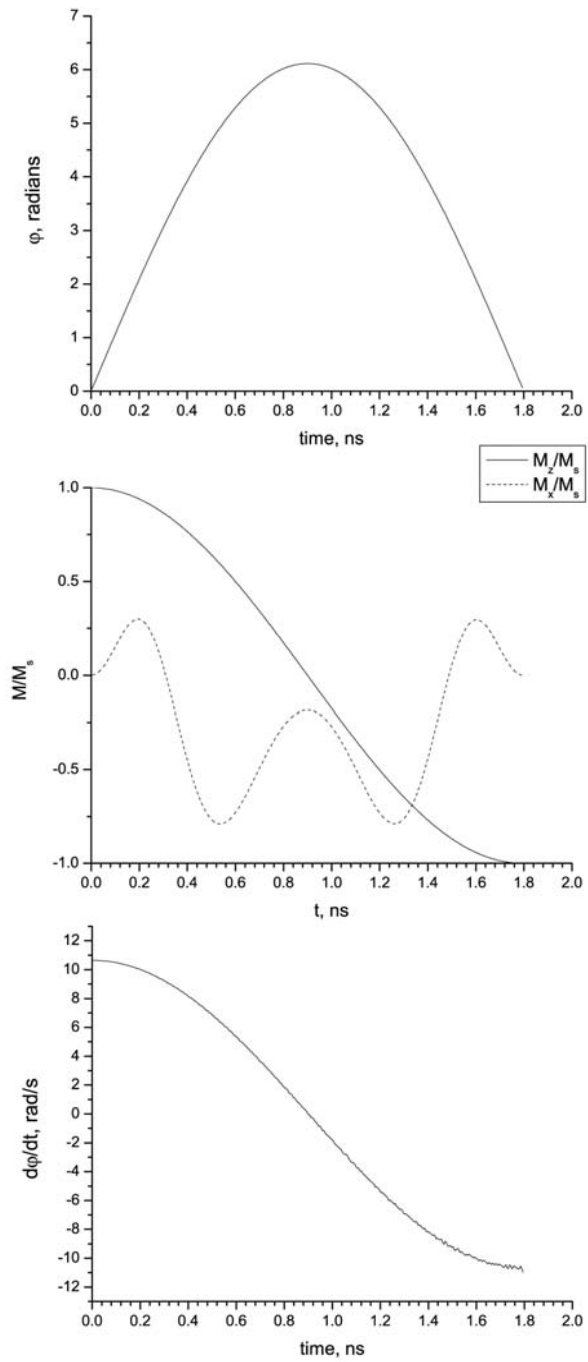

**Figure 3. Values of phase, $M_z$, $M_x$ and frequency as a function of time for**

$H_1 = 100\,\mathrm{Oe}$.

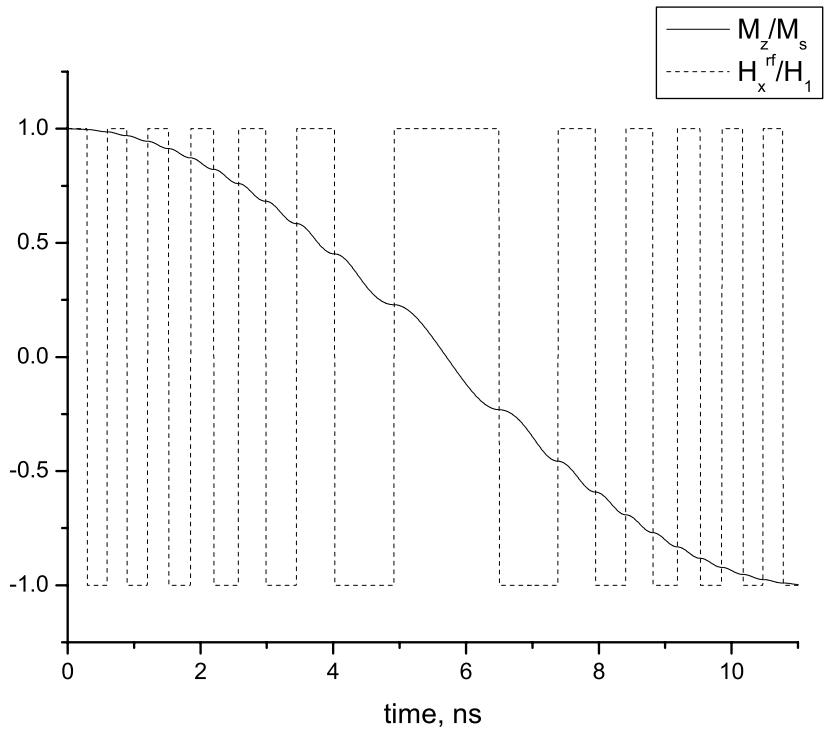

**Figure 4. Values of M$_z$ and the r.f. field as a function of time for** $H_1 = 25\,\text{Oe}$ **applied only along the x direction.**